\documentclass[footinbib, twocolumn, showpacs, pre, aps, superscriptaddress, floatfix]{revtex4-2}

\usepackage{amsmath, amssymb, amsfonts, amsthm}
\usepackage{mathtools}
\usepackage[dvipdfmx]{graphics}

\usepackage{xcolor}

\begingroup\makeatletter\catcode`,=\active\global\let\breqn@comma,\protected\gdef,{\ifmmode\expandafter\breqn@comma\else\expandafter\active@comma\fi}\endgroup

\begin{document}
\title{Pattern dynamics of the nonreciprocal Swift-Hohenberg model}

\author{Yuta Tateyama}
\affiliation{Department of Physics, Graduate School of Science, Chiba University, Chiba 263-8522, Japan}

\author{Hiroaki Ito}
\affiliation{Department of Physics, Graduate School of Science, Chiba University, Chiba 263-8522, Japan}

\author{Shigeyuki Komura}
\affiliation{Wenzhou Institute, University of Chinese Academy of Sciences, Wenzhou, Zhejiang 325001, China}
\affiliation{Oujiang Laboratory, Wenzhou, Zhejiang 325000, China} 
\affiliation{Department of Chemistry, Graduate School of Science, Tokyo Metropolitan University, Tokyo 192-0397, Japan}

\author{Hiroyuki Kitahata}\email{Corresponding author: kitahata@chiba-u.jp}
\affiliation{Department of Physics, Graduate School of Science, Chiba University, Chiba 263-8522, Japan}

\date{\today}

\begin{abstract}
    We investigate the pattern dynamics of the one-dimensional nonreciprocal Swift-Hohenberg model.
    Characteristic spatiotemporal patterns such as disordered, aligned, swap, chiral-swap, and chiral phases emerge depending on the parameters.
    We classify the characteristic spatiotemporal patterns obtained in numerical simulation by focusing on the spatiotemporal Fourier spectrum of the order parameters.
    We derive a reduced dynamical system by using the spatial Fourier series expansion.
    We analyze the bifurcation structure around the fixed points corresponding to the aligned and chiral phases, and explain the transitions between them.
    The disordered phase is destabilized either to the aligned phase by the Turing bifurcation or to the chiral phase by the wave bifurcation, while the aligned phase and the chiral phase are connected by the pitchfork bifurcation.
\end{abstract}
\maketitle

\section{Introduction}
In recent years, the concept of nonreciprocity has attracted attention in the field of nonequilibrium physics, including pattern formation and active matter since nonreciprocity leads to novel nonequilibrium phenomena~\cite{fruchart2021non}.
Non-reciprocity is often characterized by the violation of Newton's third law by effective interactions~\cite{ivlev2015statistical, ishikawa2022pairing}.
For example, microorganisms interact nonreciprocally through complex signal transduction mechanisms, which is known as quorum sensing and chemotaxis~\cite{theveneau2013chase}.
Moreover, nonreciprocity can induce new types of active phase separation phenomena~\cite{agudo2019active, zhang2021active, wittkowski2014scalar, kant2024bulk}.

Recently, phase separation dynamics with nonreciprocity due to nonequilibrium chemical potentials have been studied by extending the gradient dynamics model with a free energy functional~\cite{kant2024bulk, you2020nonreciprocity, saha2020scalar, frohoff2023stationary}.
Well-known continuum models describing phase separation are, for example, a nonconserved model represented by the Allen-Cahn equation~\cite{AC} and a conserved model represented by the Cahn-Hilliard equation~\cite{CH}.
These equations can be derived by considering gradient dynamics with proper free energy functionals~\cite{thiele2016gradient}.
Some of the authors have examined the nonreciprocal Allen-Cahn (NRAC) model, which extends the Allen-Cahn equation to a nonreciprocal system, by focusing on topological defects in two-dimensional (2D) systems~\cite{liu2023non}.
There have been several studies on the nonreciprocal Cahn-Hilliard (NRCH) model~\cite{you2020nonreciprocity, saha2024phase, saha2020scalar, frohoff2023stationary, rana2024defect, PhysRevE.103.042602, frohoff2023nonreciprocal, saha2022effervescent, suchanek2023entropy, greve2024maxwell, frohoff2021localized, brauns2024nonreciprocal, kant2024bulk}.
In the 2D NRCH model, nonreciprocity can lead to the emergence of characteristic patterns, such as targets and spirals, in which parity and time-reversal symmetry can be broken~\cite{rana2024defect}.
Generally, the introduction of nonreciprocity into the gradient dynamics can lead to characteristic spatiotemporal dynamics, such as sustained oscillation and wave propagation.

The Swift-Hohenberg (SH) equation was originally introduced as a phenomenological model to describe the formation of roll patterns in thermal convection~\cite{swift1977hydrodynamic}.
It can be derived from a nonconserved gradient dynamics model with a free energy functional.
The SH equation has parameters associated with a convective instability and a characteristic wavenumber of emerging spatial patterns~\cite{cross1993pattern}.
Several studies have been reported on the 1D coupled SH equations.
Schuler \textit{et al.}\ studied the coupled SH equations, in which two equations with different characteristic wavenumbers are coupled with nonreciprocal interactions~\cite{schuler2014spatio}.
Becker \textit{et al.}\ reexamined the coupled SH equations with nonreciprocal interaction from the viewpoint of Turing and wave bifurcations~\cite{becker2018local}.
Later Fruchart \textit{et al.}\ reported the qualitative pattern dynamics of the coupled SH equations with both reciprocal and nonreciprocal linear interactions~\cite{fruchart2021non}.
Although they reported on characteristic spatiotemporal patterns, the detailed bifurcation structure remains unexplored.

In this study we focus on the bifurcation structure of the nonreciprocal Swift-Hohenberg (NRSH) model.
We classify characteristic spatiotemporal phases obtained in numerical simulation and construct a phase diagram by using the spatiotemporal Fourier spectrum.
Since the spatial Fourier modes contributing to the spatiotemporal dynamics are limited, we can reduce the dynamics to a low-dimensional dynamical system by focusing only on the dominant spatial Fourier modes.
We demonstrate that the bifurcation structures of the pattern dynamics of the original partial differential equations can be largely understood through the analysis of the reduced model.

Our work is constructed as follows.
In Sec.~\ref{sec:model} we introduce the NRSH model and show five characteristic spatiotemporal patterns.
In Sec.~\ref{sec:identification-patterns} we construct the phase diagram based on the characteristics of the spatiotemporal Fourier spectrum of the order parameters.
In Sec.~\ref{sec:bifurcation-analysis} we perform a bifurcation analysis based on the amplitude equation derived from the spatial Fourier modes.
In Sec.~\ref{sec:discussion} we discuss the similarities between the NRSH model, the NRCH model, and the complex SH equation from the viewpoint of the amplitude equation.
Finally, we conclude our work in Sec.~\ref{sec:conclusion}.

\section{Non-reciprocal Swift-Hohenberg model}\label{sec:model}
\subsection{Governing equations}
We consider the 1D NRSH model with nonconserved real order parameters $\phi(x, t)$ and $\psi(x, t)$ described as
\begin{align}
    \dot{\phi} &= \left[\varepsilon - \left(1 + \partial_x^2\right)^2\right] \phi - \phi^3 - (\chi + \alpha) \psi,
    \label{eq:NRSH-phi}
    \\
    \dot{\psi} &= \left[\varepsilon - \left(1 + \partial_x^2\right)^2\right] \psi - \psi^3 - (\chi - \alpha) \phi.
    \label{eq:NRSH-psi}
\end{align}
Here the dot denotes the partial derivative with respect to the time $t$, $\partial_x$ represents the partial derivative with respect to the spatial coordinate $x$, $\varepsilon$ is a parameter responsible for the destabilization of spatially homogeneous states, and $\chi$ and $\alpha$ are the reciprocal and nonreciprocal interaction coefficients, respectively.
The nonreciprocal interaction term with the coefficient $\alpha$ cannot be derived from the derivative of any free energy functional, and it can originate from a nonequilibrium chemical potential~\cite{saha2020scalar}.
In our model the characteristic wavenumber of destabilization is common to $\phi$ and $\psi$.
If $(\phi, \psi)$ is a solution, $(-\phi, -\psi)$ is also a solution.
Changing the sign of $\alpha$ makes the equation equivalent by exchanging $\phi$ and $\psi$.
Changing the sign of $\chi$ makes the equation equivalent by exchanging $\phi$ and $-\psi$.
Therefore, considering only the region where $\chi \geq 0$ and $\alpha \geq 0$ is sufficient.

\subsection{Characteristic spatiotemporal patterns}\label{subsec:characteristic-patterns}
\begin{figure*}[tbp]
    \centering
    \includegraphics[width=17.6cm]{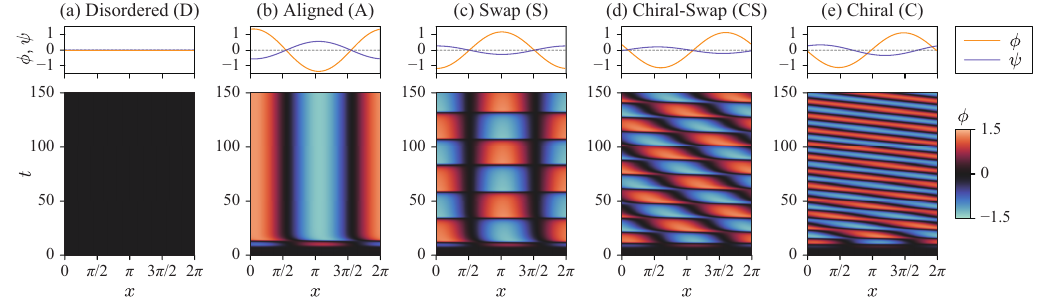}
    \caption{
        Five characteristic spatiotemporal patterns obtained by numerical simulation.
        The upper plots show the spatial distribution of $\phi$ (orange) and $\psi$ (purple) at $t = 150$.
        The lower diagrams show the spatiotemporal plots of $\phi$.
        The reciprocal interaction parameter is fixed to $\chi = 1$.
        (a) Disordered phase (D phase) when $(\varepsilon, \alpha) = (-1, 1.11)$.
        (b) Aligned phase (A phase) when $(\varepsilon, \alpha) = (0.5, 1.11)$.
        (c) Swap phase (S phase) when $(\varepsilon, \alpha) = (0.5, 1.13)$.
        (d) Chiral-swap phase (CS phase) when $(\varepsilon, \alpha) = (0.5, 1.15)$.
        (e) Chiral phase (C phase) when $(\varepsilon, \alpha) = (0.5, 1.19)$.
    }
    \label{fig:spatiotemporal-pattern-L2pi}
\end{figure*}

Here we show five characteristic spatiotemporal patterns reported in the previous study~\cite{fruchart2021non}.
We numerically solve the 1D NRSH model in Eqs.~\eqref{eq:NRSH-phi} and \eqref{eq:NRSH-psi} under periodic boundary conditions.
The system size is set to $L = 2\pi$.
The details of the numerical scheme are described in Appendix~\ref{appsec:numerical-scheme}.
The NRSH model exhibits convergence to steady spatiotemporal patterns as shown in Figs.~\ref{fig:spatiotemporal-pattern-L2pi}(a)--(e).
We fix $\chi = 1$ and vary the parameters $\varepsilon$ and $\alpha$.
We set $\varepsilon = -1$ in Fig.~\ref{fig:spatiotemporal-pattern-L2pi}(a) and $\varepsilon = 0.5$ in (b)--(e).

Figure \ref{fig:spatiotemporal-pattern-L2pi}(a) shows the disordered phase (D phase), where no spatial structure appears.
The aligned phase (A phase) is shown in Fig.~\ref{fig:spatiotemporal-pattern-L2pi}(b), where a spatially periodic wave remains stationary.
In Fig.~\ref{fig:spatiotemporal-pattern-L2pi}(c), the swap phase (S phase) is shown, which is characterized by a spatially periodic wave with an oscillating amplitude.
The chiral-swap phase (CS phase) appears as shown in Fig.~\ref{fig:spatiotemporal-pattern-L2pi}(d), which exhibits both amplitude oscillations and spatial propagation in one of the directions.
The chiral phase (C phase) is characterized by a spatially periodic wave propagating at a constant speed with a constant amplitude, as shown in Fig.~\ref{fig:spatiotemporal-pattern-L2pi}(e).

\section{Identification of patterns}\label{sec:identification-patterns}
We employ Fourier spectra to quantitatively classify the five characteristic spatiotemporal patterns obtained in the numerical results.
We limit our discussion to the case of $L = 2 \pi$ to simplify the analysis.
The spatiotemporal Fourier expansion is defined by
\begin{equation}
    \phi(x, t) = \sum_{n = -N_{\mathrm{max}}^{(x)} / 2}^{N_{\mathrm{max}}^{(x)} / 2} \sum_{m = -N_{\mathrm{max}}^{(t)} / 2}^{N_{\mathrm{max}}^{(t)} / 2} e^{- i (n k_0 x - m \omega_0 t )} \tilde{\phi}_{n,m},
\end{equation}
where $k_0 = {2 \pi}/{L} = 1$ and $\omega_0 = {2 \pi}/{T_{\mathrm{max}}}$.
The time evolution was computed up to $T_{\mathrm{max}} = 2 \times 10^4$, and the data for $0 \leq t < 4 \times 10^3$ were disregarded.
The sampling numbers were $N_{\mathrm{max}}^{(x)} = 128$ and $N_{\mathrm{max}}^{(t)} = 4 \times 10^4$.

\begin{figure}[tbp]
    \centering
    \includegraphics[width=8.6cm]{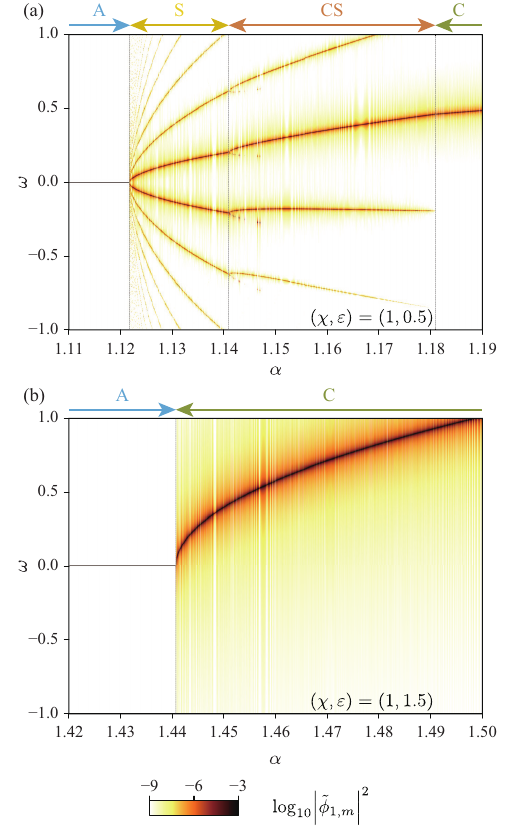}
    \caption{
        Dependence of Fourier spectra $\left|\tilde{\phi}_{1, m}\right|^2$ on $\alpha$.
        Spectra are reflected with respect to $\omega = 0$ if the maximum value of $\left|\tilde{\phi}_{1, m}\right|^2$ is found at $\omega < 0$.
        The corresponding spatiotemporal phases are shown above each spectrum.
        The parameter is fixed to $\chi = 1$, while (a) $\varepsilon = 0.5$ and (b) $\varepsilon = 1.5$.
    }
    \label{fig:fourier-Spectrum}
\end{figure}

The Fourier power spectrum $\left|\tilde{\phi}_{1, m}\right|^2$ is shown in Figs.~\ref{fig:fourier-Spectrum}(a) and (b) for the cases of $\varepsilon = 0.5$ and $\varepsilon = 1.5$, respectively, when $\chi = 1$.
As $\alpha$ increases, the behavior of the peak splitting is different between Figs.~\ref{fig:fourier-Spectrum}(a) and (b).
Hereafter we denote $\omega = m \omega_0$.
Because of the spatial inversion symmetry inherent in the 1D NRSH model, the spectrum is reflected with respect to $\omega = 0$ if the maximum peak of $\left|\tilde{\phi}_{1, m}\right|^2$ is found at $\omega < 0$.
This procedure guarantees that the maximum peak is located at $\omega \geq 0$.
In the following we discuss the case of Fig.~\ref{fig:fourier-Spectrum}(a), where more abundant phases appear, and describe the criteria for determining the spatiotemporal phases.

In the region where $\alpha$ is small, there is only one peak at $\omega = 0$, which is defined as the A phase with a stationary solution.
When $\alpha \simeq 1.12$, the spectrum splits, and peaks appear at both $\omega_+ > 0$ and $\omega_- < 0$.
In particular, when $\omega_+ = -\omega_-$ and $\left|\tilde{\phi}_+\right|^2 = \left|\tilde{\phi}_-\right|^2$, the amplitudes of the leftward and rightward traveling waves are equal, which is defined as the S phase with a standing wave-like dynamics.
In the case $\alpha \gtrsim 1.12$ and $\omega_+ + \omega_- \ne 0$, both traveling wave-like oscillations and amplitude oscillations are observed, which is defined as the CS phase.
For larger $\alpha$, there is only one peak in the region of $\omega > 0$.
This represents a traveling wave with a stationary waveform, which is defined as the C phase.

\begin{figure*}[tbp]
    \centering
    \includegraphics[width=17.6cm]{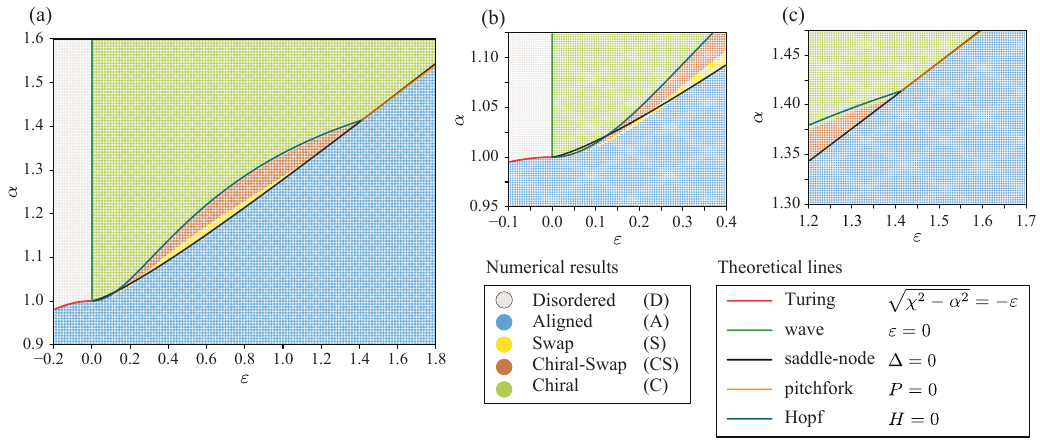}
    \caption{
        (a) Phase diagram in the $\varepsilon$-$\alpha$ plane when $L = 2\pi$ and $\chi = 1$.
        The criteria for each phase are described in Appendix~\ref{appsec:Fourier-spectrum}.
        The phase diagram obtained from numerical simulation is overlaid with theoretical lines obtained from the analysis of the reduced dynamical system in Sec.~\ref{sec:bifurcation-analysis}.
        The theoretical lines correspond to the Turing bifurcation (red line), wave bifurcation (green line), saddle-node bifurcation (black line), pitchfork bifurcation (orange line), and Hopf bifurcation (blue line).
        (b) Enlarged view near $\varepsilon = 0$ and $\alpha = 1$.
        (c) Enlarged view near $\varepsilon = \alpha = \sqrt{2}$.
    }
    \label{fig:phase-diagram-theoritical-line}
\end{figure*}

Figure~\ref{fig:phase-diagram-theoritical-line} shows the phase diagram obtained by the above classification when $\chi = 1$.
The quantitative criteria for determining the phases are described in Appendix~\ref{appsec:Fourier-spectrum}.
If we focus on the lower left region of the phase diagram in Fig.~\ref{fig:phase-diagram-theoritical-line}(a) [enlarged in Fig.~\ref{fig:phase-diagram-theoritical-line}(b)], the A phase appears even when $\varepsilon < 0$ due to the coupling interactions.
With an increase in $\varepsilon$, the D phase changes to the A phase and C phase, by Turing and wave bifurcations, respectively.

In the region with positive $\varepsilon$, an increase in $\alpha$ leads to the appearance of the A phase, S phase, CS phase, and C phase in this order in the range of $0 < \varepsilon \lesssim 1.4$, and there is a direct transition from the A phase to the C phase in the range of $\varepsilon \gtrsim 1.4$.
At the point $\alpha \simeq 1$ and $\varepsilon \simeq 0$~[see Fig.~\ref{fig:phase-diagram-theoritical-line}(b)] and the point $\alpha \simeq \varepsilon \simeq 1.4$~[see Fig.~\ref{fig:phase-diagram-theoritical-line}(c)], each phase boundary converges to a single point.
The details of bifurcation structures and the derivation of the theoretical lines are discussed in Sec.~\ref{sec:bifurcation-analysis}.

\section{Theoretical analysis}\label{sec:bifurcation-analysis}
\subsection{Linear stability analysis}
We first perform a linear stability analysis around the D phase and show that the A phase and C phase appear through Turing and wave bifurcations, respectively.
We perform the spatial Fourier series expansion to investigate the appearance of patterns with finite wavenumbers.
In the following we consider the dynamics of $\phi(x, t)$ and $\psi(x, t)$ in $0 \leq x < L$ with a periodic boundary condition.
The Fourier series expansion with respect to $x$ is described as
\begin{align}
    \phi(x, t) &= \sum_{n = -\infty}^\infty \phi_n(t) e^{-i n k_0 x},
    \label{eq:phi-Fourier}
    \\
    \psi(x, t) &= \sum_{n = -\infty}^\infty \psi_n(t) e^{-i n k_0 x}.
    \label{eq:psi-Fourier}
\end{align}
Because both $\phi$ and $\psi$ are real variables, the Fourier coefficients satisfy the relations $\phi_n = \phi^*_{-n}$ and $\psi_n = \psi^*_{-n}$, where ${}^*$ denotes the complex conjugate.
By the substitution of Eqs.~\eqref{eq:phi-Fourier} and \eqref{eq:psi-Fourier} into the NRSH model in Eqs.~\eqref{eq:NRSH-phi} and \eqref{eq:NRSH-psi}, the ordinary differential equations for the $n$-th mode Fourier components are derived as
\begin{align}
    \dot{\phi}_n &= \left[
        \varepsilon - \left(1 - n^2 k_0^2\right)^2
    \right] \phi_n - F_n(\{\phi_m\}) - (\chi + \alpha) \psi_n,
    \label{eq:NRSH-phi-n}
    \\
    \dot{\psi}_n &= \left[
        \varepsilon - \left(1 - n^2 k_0^2\right)^2
    \right] \psi_n - F_n(\{\psi_m\}) - (\chi - \alpha) \phi_n,
    \label{eq:NRSH-psi-n}
\end{align}
where the third-order nonlinear term $F_n$ is defined as
\begin{align}
    F_n(\{\phi_m\}) &= \sum_{m_1 = -\infty}^\infty \sum_{m_2 = -\infty}^\infty \phi_{m_1} \phi_{m_2} \phi_{n - m_1 - m_2},
    \label{eq:F_n-phi}
    \\
    F_n(\{\psi_m\}) &= \sum_{m_1 = -\infty}^\infty \sum_{m_2 = -\infty}^\infty \psi_{m_1} \psi_{m_2} \psi_{n - m_1 - m_2}.
    \label{eq:F_n-psi}
\end{align}
Since $n$ takes all integers, Eqs.~\eqref{eq:NRSH-phi-n} and \eqref{eq:NRSH-psi-n} represent a coupled system of an infinite number of ordinary differential equations for the Fourier coefficients $\{\phi_n, \psi_n\}$.

Here we consider the case with $k_0 = 1$ for simplicity.
In this case, the eigenvalues obtained from the linearized equations around the trivial solution $(\phi, \psi) = (0, 0)$ are
\begin{equation}
    \lambda_n^\pm
    = \varepsilon - \left(1 - n^2\right)^2
    \pm \sqrt{\chi^2 - \alpha^2}.
    \label{eq:trivial-eigenvalue}
\end{equation}
It should be noted that the real part of the eigenvalues is maximized when $n = \pm 1$.
When $\alpha \leq \chi$, the instability condition $\lambda_{\pm1}^+ > 0$ is satisfied if $\sqrt{\chi^2 - \alpha^2} > -\varepsilon$.
The boundary line, $\sqrt{\chi^2 - \alpha^2} = -\varepsilon$, is shown by the red line in Fig.~\ref{fig:phase-diagram-theoritical-line}.
In this case, the eigenvalues do not have the imaginary part, which corresponds to the Turing instability.
When $\alpha > \chi$, the real part of $\lambda_{\pm1}^+$ is $\varepsilon$, and thus the D phase is destabilized when $\varepsilon > 0$, leading to the appearance of the C phase.
Moreover, $\lambda_{\pm1}^+$ has an imaginary part $\pm i \sqrt{\alpha^2 - \chi^2}$, which implies the wave instability.
The boundary line, $\varepsilon = 0$, is shown by the green line in Fig.~\ref{fig:phase-diagram-theoritical-line}.

\subsection{Reduced ordinary differential equation for Fourier coefficients}
Next, we consider the dominant unstable modes $n = \pm 1$ and ignore the other modes in the analysis.
Under this approximation, the possible three combinations of $(m_1, m_2, 1 - m_1 - m_2)$ are $(1, 1, - 1)$, $(1, - 1, 1)$, and $(- 1, 1, 1)$ in Eqs.~\eqref{eq:F_n-phi} and \eqref{eq:F_n-psi}.
The resulting expressions for the nonlinear terms $F_1(\left\{\phi_m\right\})$ and $F_1(\{\psi_m\})$ are approximated as $3 |\phi_1|^2 \phi_1$ and $3 |\psi_1|^2 \psi_1$, respectively.
Consequently, we obtain a system of coupled ordinary differential equations for the complex order parameters $\phi_1$ and $\psi_1$ as
\begin{align}
    \dot{\phi}_1 &= \varepsilon \phi_1 - 3 |\phi_1|^2 \phi_1- (\chi + \alpha) \psi_1,
    \label{eq:phi1}
    \\
    \dot{\psi}_1 &= \varepsilon \psi_1 - 3 |\psi_1|^2 \psi_1- (\chi - \alpha) \phi_1.
    \label{eq:psi1}
\end{align}

We further reduce the degrees of freedom by focusing on the phase translational symmetry of Eqs.~\eqref{eq:phi1} and \eqref{eq:psi1}, which originates from the spatial translational symmetry.
We set $\phi_1 = \rho_1 e^{i \theta_1}$ and $\psi_1 = \rho_2 e^{i \theta_2}$, where $\rho_1$ and $\rho_2$ are non-negative real values, while $\theta_1$ and $\theta_2$ are real values.
Furthermore, by defining the phase difference as $\delta = \theta_2 - \theta_1$ and by substituting them into Eqs.~\eqref{eq:phi1} and \eqref{eq:psi1}, we obtain a three-variable dynamical system as
\begin{align}
    \dot{\rho}_1 &= \varepsilon \rho_1 - 3 \rho_1^3 - (\chi + \alpha) \rho_2 \cos \delta,
    \label{eq:rho1}
    \\
    \dot{\rho}_2 &= \varepsilon \rho_2 - 3 \rho_2^3 - (\chi - \alpha) \rho_1 \cos \delta,
    \label{eq:rho2}
    \\
    \dot{\delta} &= \left[
        (\chi - \alpha) \frac{\rho_1}{\rho_2}
        + (\chi + \alpha) \frac{\rho_2}{\rho_1}
    \right] \sin \delta.
    \label{eq:delta}
\end{align}
Note that $\dot{\theta}_1$ and $\dot{\theta}_2$ obey
\begin{align}
    \dot{\theta}_1 &= - (\chi + \alpha) \frac{\rho_2}{\rho_1} \sin \delta,
    \\
    \dot{\theta}_2 &= (\chi - \alpha) \frac{\rho_1}{\rho_2} \sin \delta,
\end{align}
and thus $\dot{\theta}_1$ and $\dot{\theta}_2$ do not necessarily become zero even if $\dot{\delta} = 0$.

\subsection{Fixed points of the reduced system and corresponding spatiotemporal patterns}
We consider the fixed points $(\rho_1^{(0)}, \rho_2^{(0)}, \delta^{(0)})$ of the reduced three-variable dynamical system in Eqs.~\eqref{eq:rho1}, \eqref{eq:rho2} and \eqref{eq:delta} by setting $\dot{\rho}_1 = \dot{\rho}_2 = \dot{\delta} = 0$.
Regarding $\dot{\delta} = 0$ in Eq.~\eqref{eq:delta}, either $\sin\delta^{(0)} = 0$ or $(\chi - \alpha) \rho_1^{(0)} / \rho_2^{(0)} + (\chi + \alpha) \rho_2^{(0)} / \rho_1^{(0)} = 0$ is necessary.
As will be seen in the subsequent analysis, the solutions for the former equation correspond to the A phase, and the ones for the latter equation correspond to the C phase.

First, we consider the fixed point with $\sin \delta^{(0)} = 0$.
We substitute Eq.~\eqref{eq:rho1} with $\dot{\rho}_1 = 0$ into Eq.~\eqref{eq:rho2} with $\dot{\rho}_2 = 0$ and define $\zeta$ as
\begin{equation}
    \zeta = \frac{\varepsilon - 3 \left(\rho_1^{(0)}\right)^2}{\chi + \alpha}.
\end{equation}
Then, we obtain a quartic equation of $\zeta$~\footnote{$\zeta$ corresponds to the solution $\rho_1^{(0)}$ if and only if $\varepsilon - (\chi + \alpha) \zeta > 0$ holds.
$\phi_1$, $\psi_1$, and $\zeta$ should have following relations: $\psi_1 = \zeta \phi_1$ and $\arg \psi_1 = \arg \phi_1$ when $\zeta > 0$, and $\arg \psi_1 = \arg \phi_1 + \pi$ when $\zeta < 0$.
The case where $\zeta = 0$ and $\psi_1 = 0$ is the only case where the phase factor of $\psi_1$ is undefined, and this occurs only when $\chi = \alpha$.
When $\chi = \alpha$, $|\phi_1|^2 = \varepsilon / 3$ and $\zeta = 0$, that is, $|\phi_1| = \sqrt{\varepsilon / 3}$ and $\psi_1 = 0$ are fixed points.
In this case, $\phi_1 \ne 0$.
Therefore, it is not the D phase, and the phase difference $\delta$ cannot be defined because $\psi_1$ is zero.
Such a singularity is known as an exceptional point.} as
\begin{equation}
    (\chi + \alpha) \zeta^4 - \varepsilon \zeta^3 + \varepsilon \zeta - (\chi - \alpha) = 0.
    \label{eq:zeta-4}
\end{equation}
The discriminant $\Delta$ of the above quartic equation is given by
\begin{align}
    \frac{\Delta}{4} ={} &
    64 \alpha^{6} - 192 \alpha^{4} \chi^{2} + 48 \alpha^{4} \varepsilon^{2} + 192 \alpha^{2} \chi^{4} - 96 \alpha^{2} \chi^{2} \varepsilon^{2}
    \notag
    \\
    &- 15 \alpha^{2} \varepsilon^{4} - 64 \chi^{6} + 48 \chi^{4} \varepsilon^{2} - 12 \chi^{2} \varepsilon^{4} + \varepsilon^{6}.
    \label{eq:Discriminant}
\end{align}

\begin{figure}[tbp]
    \centering
    \includegraphics[width=8.6cm]{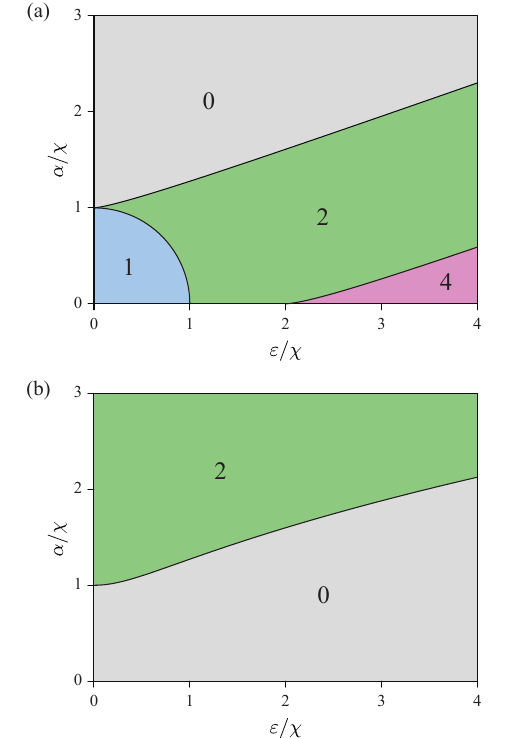}
    \caption{
        Number of fixed points in the reduced dynamical system for (a) the A phase and (b) the C phase.
    }
    \label{fig:exsistence-condition-FixedPoints}
\end{figure}

The number of fixed points changes when the sign of $\Delta$ changes.
Therefore, $\Delta = 0$ gives the saddle-node bifurcation line, shown by the black line in Fig.~\ref{fig:phase-diagram-theoritical-line}.
Considering the original spatiotemporal variables, one can recover as $\phi(x, t) = 2\rho_1^{(0)} \cos(x + \theta_0)$ and $\psi(x, t) = \pm 2\rho_2^{(0)} \cos(x + \theta_0)$, which correspond to the A phase.
Here $\theta_0$ is an arbitrary constant due to the spatial translational symmetry.
Moreover, by examining the quartic equation of $\zeta$ in detail, we find that the number of fixed points changes depending on the signs of $\Delta$ and $(\varepsilon / \chi)^2 + (\alpha / \chi)^2 - 1$.
In Fig.~\ref{fig:exsistence-condition-FixedPoints}(a), we show the number of fixed points in the $(\varepsilon / \chi)$-$(\alpha / \chi)$ plane.

Next, we consider the case when $(\chi - \alpha) \rho_1^{(0)} / \rho_2^{(0)} + (\chi + \alpha) \rho_2^{(0)} / \rho_1^{(0)} = 0$ holds, which is the other sufficient condition for $\dot{\delta} = 0$.
In this case, the fixed points are explicitly given by
\begin{align}
    \rho_1^{(0)} &= \sqrt{
        \frac{\varepsilon}{3 \alpha} (\alpha + \chi)
    },
    \\
    \rho_2^{(0)} &= \sqrt{
        \frac{\varepsilon}{3 \alpha} (\alpha - \chi)
    },
    \\
    \cos\delta^{(0)} &= -\frac{\chi \varepsilon}{\alpha \sqrt{\alpha^2 - \chi^2}}.
\end{align}
The existence of this fixed point is equivalent to the conditions $\varepsilon > 0$, $\alpha > \chi$, and $P \geq 0$, where $P$ is defined as
\begin{equation}
    P = \alpha^4 - \alpha^2\chi^2 - \chi^2\varepsilon^2.
\end{equation}
As shown in the next subsection, the boundary line $P = 0$ corresponds to the pitchfork bifurcation, which is shown by the orange line in Fig.~\ref{fig:phase-diagram-theoritical-line}.
The fixed point in the reduced three-variable dynamical system corresponds to the solution $\dot{\theta}_1 = \dot{\theta}_2$ for the four-variable dynamical system.
Here,
\begin{equation}
    \left|\dot{\theta}_1\right| = \left|\dot{\theta}_2\right| = \frac{\sqrt{P}}{\alpha} = \sqrt{\alpha^2 - \chi^2 - \frac{\chi^2 \varepsilon^2}{\alpha^2}},
\end{equation}
where the signs of $\dot{\theta}_1$ and $\dot{\theta}_2$ indicate the direction of the wave propagation in real space.
Therefore, $\phi(x, t)$ and $\psi(x, t)$ propagate with a constant phase difference $\delta^{(0)}$ and a phase velocity $\Omega_0 = \sqrt{P} / \alpha$ in the original spatiotemporal variables, $\phi(x, t) = 2\rho_1^{(0)} \cos(x \mp \Omega_0 t + \theta_1^{(0)})$ and $\psi(x, t) = 2\rho_2^{(0)} \cos(x \mp \Omega_0 t + \theta_1^{(0)} + \delta^{(0)})$, which corresponds to the C phase.
The different signs $\mp$ represent rightward and leftward propagating waves, respectively.

In Fig.~\ref{fig:exsistence-condition-FixedPoints}(b), we show the number of fixed points for the C phase in the $(\alpha / \chi)$-$(\varepsilon / \chi)$ plane.
The stability condition for the fixed points for the C phase is derived by using the Routh-Hurwitz stability criterion~\cite{gradshteyn2014table}:
\begin{equation}
    H = \alpha^4 - \alpha^2 \chi^2 + 2 \alpha^2 \varepsilon^2 - 5 \chi^2 \varepsilon^2 > 0.
\end{equation}
Upon examining the eigenvalues of the Jacobi matrix around the fixed point for the C phase on the boundary line $H = 0$, we obtain one negative eigenvalue and two purely imaginary eigenvalues.
This suggests that the boundary $H = 0$ corresponds to the Hopf bifurcation line, which is shown by the blue line in Fig.~\ref{fig:phase-diagram-theoritical-line}.

As shown in Fig.~\ref{fig:phase-diagram-theoritical-line}(a), the five theoretical lines of Turing, wave, saddle-node, pitchfork, and Hopf bifurcations reproduce well the phase boundaries of the numerical results.
However, in the region of $\varepsilon \simeq 0$ and $\alpha \simeq 1$ in Fig.~\ref{fig:phase-diagram-theoritical-line}(b), the S phase exists below both the saddle-node (black) and Hopf (blue) bifurcation lines.
This will be explained by the existence of the global bifurcation, as discussed in the next subsection.

\subsection{Connection of branches and bifurcation structures}\label{subsec:branch-connection}
\begin{figure}[tbp]
    \centering
    \includegraphics[width=8.6cm]{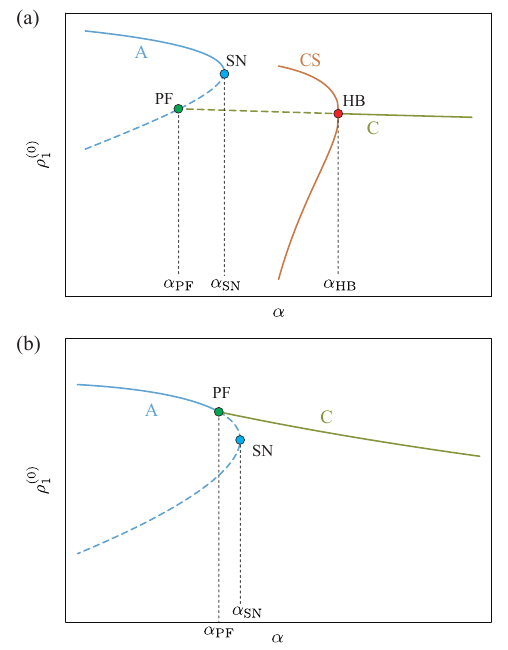}
    \caption{
        Schematic diagrams of the bifurcation structure of the fixed points for the A phase (blue line) and C phase (green line) when (a) $\varepsilon / \chi < \sqrt{2}$ and (b) $\varepsilon / \chi > \sqrt{2}$.
        The maximum and minimum values of the periodic orbits corresponding to the CS phase generated by the Hopf bifurcation are shown by the orange line.
        Solid and dashed lines represent stable and unstable fixed points, respectively.
        PF: pitchfork bifurcation point, SN: saddle-node bifurcation point, HB: Hopf bifurcation point.
    }
    \label{fig:branch-connection}
\end{figure}

In the previous subsection, we investigated the existence and stability conditions in the reduced dynamical system for the spatial Fourier modes.
The fixed points for the A phase and C phase are obtained from the conditions $\sin \delta^{(0)} = 0$ and $(\chi - \alpha) \rho_1^{(0)} / \rho_2^{(0)} + (\chi + \alpha) \rho_2^{(0)} / \rho_1^{(0)} = 0$, respectively.
The fixed points for the C phase also satisfy $\sin \delta^{(0)} = 0$ if $P = 0$, leading to an intersection of the branches of the A phase and C phase at the pitchfork bifurcation point.
The bifurcation structure depends on the ratio $\varepsilon / \chi$, as schematically shown in Figs.~\ref{fig:branch-connection}(a) and (b).

When $\varepsilon / \chi < \sqrt{2}$ in Fig.~\ref{fig:branch-connection}(a), there are three characteristic values of $\alpha$: the pitchfork bifurcation point $\alpha_{\mathrm{PF}}$, the saddle-node bifurcation point $\alpha_{\mathrm{SN}}$, and the Hopf bifurcation point $\alpha_{\mathrm{HB}}$, where $\alpha_{\mathrm{PF}} < \alpha_{\mathrm{SN}} < \alpha_{\mathrm{HB}}$.
These points are characterized by $P = 0$, $\Delta = 0$, and $H = 0$, respectively.
For $0 \leq \alpha < \alpha_{\mathrm{PF}}$, there are two fixed points for the A phase, where one is stable and the other is unstable.
At the pitchfork bifurcation point $\alpha = \alpha_{\mathrm{PF}}$, two unstable fixed points for the C phase emerge from the unstable fixed point for the A phase.
Note that the value of $\delta^{(0)}$ for the fixed points corresponding to the C phase are symmetric with respect to $\pi$,  reflecting the symmetry of $\cos\delta$ around $\delta = \pi$.
Hence, for $\alpha_{\mathrm{PF}} < \alpha < \alpha_{\mathrm{SN}}$, there are four fixed points; two of which are stable and unstable fixed points for the A phase, and the other two are unstable fixed points for the C phase.
At the saddle-node bifurcation point $\alpha = \alpha_{\mathrm{SN}}$, the stable and unstable fixed points for the A phase merge and disappear.
For $\alpha_{\mathrm{SN}} < \alpha < \alpha_{\mathrm{HB}}$, there are two fixed points, both of which are unstable, corresponding to the C phase.
At the Hopf bifurcation point $\alpha = \alpha_{\mathrm{HB}}$, the two fixed points for the C phase change stability; they are stable for $\alpha > \alpha_{\mathrm{HB}}$ and unstable for $\alpha < \alpha_{\mathrm{HB}}$.
Moreover, for $\alpha < \alpha_{\mathrm{HB}}$, two stable closed orbits, which correspond to the CS phase, are generated from each Hopf bifurcation point. 
For $\alpha > \alpha_{\mathrm{HB}}$, there are two fixed points for the C phase, both of which are stable.

When $\varepsilon / \chi > \sqrt{2}$ in Fig.~\ref{fig:branch-connection}(b), there are two characteristic values of $\alpha$; the pitchfork bifurcation point $\alpha_{\mathrm{PF}}$ and the saddle-node bifurcation point $\alpha_{\mathrm{SN}}$, where $\alpha_{\mathrm{PF}} < \alpha_{\mathrm{SN}}$.
For $\alpha < \alpha_{\mathrm{PF}}$, there are two fixed points for the A phase, one of which is stable and the other is unstable.
At the pitchfork bifurcation point $\alpha = \alpha_{\mathrm{PF}}$, two stable fixed points corresponding to the C phase are generated from the stable fixed point corresponding to the A phase.
Through this pitchfork bifurcation, the stable fixed point for the A phase becomes unstable, and both fixed points corresponding to the A phase are unstable for $\alpha_{\mathrm{PF}} < \alpha < \alpha_{\mathrm{SN}}$.
For $\alpha_{\mathrm{PF}} < \alpha < \alpha_{\mathrm{SN}}$, there are four fixed points; two of which are unstable fixed points corresponding to the A phase, and the other two are stable fixed points for the C phase.
At the saddle-node bifurcation point $\alpha = \alpha_{\mathrm{SN}}$, the two unstable fixed points for the A phase merge and disappear.
For $\alpha > \alpha_{\mathrm{SN}}$, there are two fixed points corresponding to the C phase, and both of them are stable.

When $\varepsilon / \chi = \sqrt{2}$, the pitchfork, saddle-node, and Hopf bifurcations degenerate at $\alpha_{\mathrm{PF}} = \alpha_{\mathrm{SN}} = \alpha_{\mathrm{HB}} = \sqrt{2} \chi$, suggesting a high-codimensional bifurcation~\cite{arnold2012geometrical, hale2002dynamics}.
Similarly, at the point $\alpha / \chi = 1$ and $\varepsilon / \chi = 0$, the wave bifurcation point and the Turing bifurcation point coincide.

According to numerical calculations of the reduced dynamical system for the complex order parameters Eqs.~\eqref{eq:phi1} and \eqref{eq:psi1} (data not shown), the S phase appears from the A phase by a global bifurcation.
For example, when $\varepsilon / \chi$ is slightly less than $\sqrt{2}$, a periodic orbit corresponding to the S phase emerges through an infinite period bifurcation at $\alpha = \alpha_{\mathrm{SN}}$.
Furthermore, when $\varepsilon / \chi$ is small, the S phase is suggested to appear through a heteroclinic bifurcation around the unstable fixed points for the A phase.
The details of these global bifurcations are left for our future research.

\section{Discussion}\label{sec:discussion}
Several studies have reported on the 1D complex SH equation~\cite{sakaguchi1997standing, sakaguchi1998localized, gelens2011traveling, khairudin2016stability}.
The typical form of the complex SH equation is written as
\begin{equation}
    \dot{w} = \left[c_1 - c_2 \left(1 + \partial_x^2\right)^2\right] w - c_3 |w|^2 w,
    \label{eq:complex-SH}
\end{equation}
where $w$ is a complex order parameter, and $c_1$, $c_2$, and $c_3$ are complex parameters.
The complex SH equation no longer has gradient dynamics when the parameters $c_1$, $c_2$, and $c_3$ are complex.
There are reports of traveling wave patterns similar to those in the 1D NRSH model~\cite{sakaguchi1997standing, gelens2011traveling}.
However, in the complex SH equation with translational symmetry of the phase factor for $w$, the reciprocal interactions corresponding to $- \chi \psi$ and $- \chi \phi$ in Eqs.~\eqref{eq:NRSH-phi} and \eqref{eq:NRSH-psi} cannot be represented even if the coefficients are complex.
Although the solutions corresponding to the A phase and C phase exist in the complex SH equation, patterns with amplitude oscillations, such as the S phase and CS phase, do not appear.
Reciprocal interactions in the NRSH model break the translational symmetry of the phase factor of $w$ and can generate patterns with amplitude oscillations.
To include reciprocal interactions in the complex SH equation, it is necessary to introduce terms that break the translational symmetry of the phase factor, such as $- i \chi w^*$, in Eq.~\eqref{eq:complex-SH}.

In the 2D complex SH equation, traveling domain wall structures and spiral structures were observed~\cite{aranson1995domain}.
Recently, an experimental study reported self-integrated atomic quantum wires and junctions of a Mott semiconductor~\cite{asaba2023growth}.
This study revealed the emergence of unique spiral patterns in reaction-diffusion systems.
To provide a theoretical analysis of these observations, a model using the 2D complex SH equation was employed.
The numerical calculations demonstrated the generation and progression of stripe patterns originating from the initial spiral structure~\cite{asaba2023growth}.
The stripe patterns that propagate with a finite characteristic wavenumber correspond to the C phase in our 1D NRSH model.
In 2D systems, topological defects appear, and the spatiotemporal pattern dynamics become more complex.

In the 1D NRSH model, we derived the reduced model by focusing on the Fourier modes around the characteristic wavenumber.
Here we discuss the relationship between the 1D NRSH model and the 1D NRCH model.
In the 1D NRCH model, amplitude equations can be obtained by a reduction around the characteristic wavenumber~\cite{you2020nonreciprocity}.
By considering the zeroth Fourier mode as an appropriate parameter, a set of amplitude equations that is completely equivalent to that in the NRSH model can be obtained.
This suggests that the spatiotemporal pattern dynamics of the NRSH and NRCH models are equivalent when the system size is sufficiently small, and only the modes around the characteristic wavenumber play an essential role.

\begin{figure*}[tbp]
    \centering
    \includegraphics[width=17.6cm]{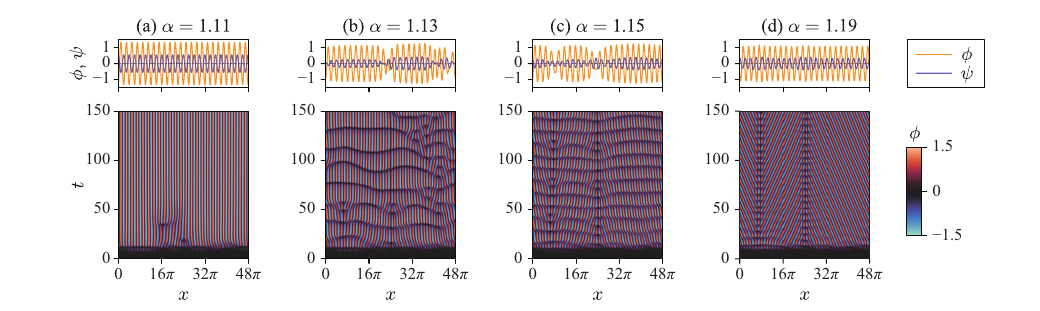}
    \caption{
        Characteristic spatiotemporal patterns observed by numerical simulation in the 1D NRSH model with $L = 48 \pi$.
        The upper plots show the spatial distribution of $\phi$ and $\psi$ at $t = 150$, where orange and purple lines represent $\phi$ and $\psi$, respectively.
        The lower diagrams show the spatiotemporal plots of $\phi$.
        The nonreciprocal interaction parameter $\alpha$ in each phase is varied.
        (a) $\alpha = 1.11$, (b) $\alpha = 1.13$, (c) $\alpha = 1.15$, and (d) $\alpha = 1.19$.
        The other parameters are fixed to $\varepsilon = 0.5$ and $\chi = 1$.
    }
    \label{fig:spatiotemporal-pattern-L48pi}
\end{figure*}

Differences between the NRSH and NRCH models appear through the coupling with other wavenumber modes, such as the exsistence of oscillations in the zeroth Fourier mode and long-wavelength instabilities known as the Eckhaus instability~\cite{nishiura2002far}, or in the systems with higher spatial dimensions.
When numerical calculations is performed on the NRSH model with a large system size with $L = 48 \pi$, long-wavelength instabilities is observed as shown in Fig.~\ref{fig:spatiotemporal-pattern-L48pi}.
The spatial pattern of the S phase slips to left and right due to the coupling with long-wavelength modes, similar to the CS phase.
In the CS phase and C phase, sinks and sources appear, and the advancing waves move towards or away from them.

\section{Conclusion}\label{sec:conclusion}
We investigated the pattern dynamics and phase transitions of the 1D NRSH model through numerical calculations and theoretical analyses.
Numerical simulations revealed that the spatially periodic wave (A phase) appears when nonreciprocity $\alpha$ is sufficiently small, while the traveling wave with a constant amplitude (C phase) appears when $\alpha$ is large.
Depending on the reciprocal parameter $\chi$ and the destabilization parameter $\varepsilon$, the S phase and CS phase appear.
By considering the Fourier modes corresponding to the characteristic wavenumber, we derived a reduced dynamical system.
Our theoretical analyses clarified the bifurcation structure of the reduced system, particularly the fixed points corresponding to the A phase and C phase.
Our analyses reproduced well the phase diagram obtained by the numerical classification of the spatiotemporal Fourier spectrum.
Our findings underscored the unique characteristics of nonreciprocal continuum systems, where transitions to dynamic phases are driven by nonreciprocity.
We also unveiled the rich bifurcation structure including S phase and CS phase, which is due to the presence of both reciprocal and nonreciprocal interactions.

\section*{Acknowledgments}
Y.T.\ acknowledges the support by the Japan Science and Technology Agency (JST), the establishment of university fellowships towards the creation of science technology innovation (No.\ JPMJFS2107).
H.I.\ acknowledges the support by the Japan Society for the Promotion of Science (JSPS) KAKENHI Grant (No.\ JP24K06972).
S.K.\ acknowledges the support by the National Natural Science Foundation of China (Nos.\ 12274098 and 12250710127) and the startup grant of Wenzhou Institute, University of Chinese Academy of Sciences (No.\ WIUCASQD2021041).
H.K.\ acknowledges the support by JSPS KAKENHI Grant (No.\ JP21H01004).
We acknowledge the support by JSPS Core-to-Core Program ``Advanced core-to-core network for the physics of self-organizing active matter'' (No.\ JPJSCCA20230002).

\appendix
\section{Numerical scheme}\label{appsec:numerical-scheme}
The spatial grid size was $\Delta x = L / N \simeq 0.05$.
The initial condition $(\phi_0, \psi_0)$ was set by the Gaussian white noise with zero mean and variance $10^{-4}$ at each spatial point.
The 1D Laplace operator $\partial_x^2$ was approximated with second-order accuracy using a central difference scheme, and the time evolution was performed using an open-source differential equation solver employing the explicit singly diagonal implicit Runge-Kutta (ESDIRK) method~\cite{rackauckas2017differentialequations}.

\section{Criteria for determining the phases}\label{appsec:Fourier-spectrum}
We defined the criteria for determining the D phase, A phase, S phase, CS phase, and C phase.
We focused on the Fourier component $\tilde{\phi}_{1, m}$ of $\phi(x, t)$.
If the maximum value of $\left|\tilde{\phi}_{1, m}\right|^2$ as a function of $\omega$ is less than $10^{-4}$, it is determined to be the D phase.
If the maximum value of $\left|\tilde{\phi}_{1, m}\right|^2$ with respect to $\omega$ is given at $\omega = 0$, it is determined to be the A phase.
In case the maximum value is given at $\omega \ne 0$, if there is only one maximum value of $\left|\tilde{\phi}_{1, m}\right|^2$ with respect to $\omega$, or the ratio of the maximum value of $\left|\tilde{\phi}_{1, m}\right|^2$ to the next largest value of $\left|\tilde{\phi}_{1, m}\right|^2$ is less than $10^{-2}$, it is determined to be the C phase.
Let $\omega_1$ and $\omega_2$ be the frequencies of the maximum and the next largest values of $\left|\tilde{\phi}_{1, m}\right|^2$, respectively.
If $|\omega_1 + \omega_2|$ is less than $10^{- 3}$ and the difference between the maximum value of $\left|\tilde{\phi}_{1, m}\right|^2$ and the next largest value is less than $10^{- 2}$, it is determined to be the S phase.
Otherwise, it was determined to be the CS phase.

\end{document}